\documentstyle[aps,preprint]{revtex}

\begin{document}
\draft
\title{Phonon Anomalies Induced by Superconductivity }
\author{Hae-Young Kee$^{1,2}$ and C. M. Varma$^2$\\
${}^1$  Department of Physics, Rutgers University, Piscataway, NJ 08855-0849\\
${}^2$ Bell Laboratories, Lucent Technologies, Murray Hill, NJ 07974}
\maketitle

\begin{abstract}
We calculate the electronic polarizability in the superconducting state
near extremum vectors ${\vec Q}_0$ of the Fermi surface.
A pole appears in the polarizability at frequencies
$\omega$ near the superconducting gap $2\Delta$ 
which leads to a sharp peak at $\omega$ just below $2 \Delta$
in the lattice vibrations near ${\vec Q}_0$.
A second order transition to a charge density wave state
below the superconducting transition is shown to be unlikely.
The results are compared with the recent inelastic neutron
scattering measurements in compounds, $ANi_2B_2C$, $A=Y$ or $Lu$.
\end{abstract}

\newpage

Strong momentum dependence of the electronic polarizability leads to 
characteristic features in the dispersion relation of phonons in simple
metals--the Kohn anomalies.\cite{kohn}
Transition metals and compounds are particularly beset with wiggles and 
dips in their phonon dispersions.
Strong electron-phonon interactions and local field effects are responsible 
together with nesting features in the band structure for the
anomalies and for the occurrence of charge density wave transitions\cite{varma}.

It is well known that the leading order polarizability
$\Pi^n_0({\vec Q},\omega)$,
Fig. 1 (a), of an electron gas in d-dimensions for ${\vec Q}=2{\vec k_F}$
for a spherical Fermi-surface or ellipsoidal Fermi-surface
across the major and minor axes of the Fermi-surface 
in the normal state is\cite{2kpol}

\begin{equation}
{\rm Re} \Pi_0^n (2{\vec k_F},\omega) \simeq N(0)[1+(N(0)\omega)^{d-1} {\rm ln}(\frac{E_F}{\omega})].
\end{equation}

We show the existence of singularities in the polarizability
at $\omega \simeq 2\Delta$ in s-wave superconductors near the extremum vectors
of the Fermi-surface which are stronger than those of the normal state at
$\omega \rightarrow 0$.
By an extremum vector we mean ${\vec Q}_0$ such that 
$\epsilon_{{\vec Q}_0/2}$ and
 $\epsilon_{-{\vec Q}_0/2}$ are both
on the Fermi-surface and for small deviations $|{\vec k}| << |{\vec Q}_0|$

\begin{eqnarray}
\epsilon_{{\vec k}-\frac{{\vec Q}_0}{2}} &=& -v_{F\parallel} {\vec k}_\parallel+\sum_i \frac{{\vec k}^2_{i \perp}}
{2 m_{i \perp}}\nonumber\\
\epsilon_{{\vec k}+\frac{{\vec Q}_0}{2}} &=& v_{F\parallel} {\vec k}_\parallel
+\sum_i \frac{{\vec k}^2_{i \perp}}{2 m_{i \perp}}
\end{eqnarray}
Here ${\vec k}_{\parallel}$ is the component of ${\vec k}$ 
 parallel to ${\vec Q}_0$ and 
 ${\vec k}_{i \perp}$ 
are the components perpendicular to ${\vec Q}_0$ where $i$
runs over $d-1$ values.
For example, ${\vec Q}_0$ are the major and minor axes for an ellipsoidal
Fermi-surface.
${\vec Q}_0$ is also a nesting vector when $m_{i \perp}$ for one or more $i$ is
infinity.

These singularities strongly affect the phonon spectrum of superconductors
under certain conditions.
We explain the remarkable difference between the phonon spectrum
of $YNi_2B_2C$ in the superconducting state 
and that of the normal state recently observed by
Kawano et al\cite{kawano} and in $LuNi_2B_2C$ by Stassis et al\cite{stassis}.
In the normal phase of $(Y,Lu)Ni_2B_2C$ the transverse acoustic and
optic dispersion
curve in the $\Gamma Z$ direction has a strong temperature dependent dip
around ${\vec Q} = (0.5,0,8)$.
In the $Y$ compound,
below the superconducting transition temperature, the spectral weight splits
into two parts, a very sharp resolution limited line at an energy
$\simeq 2\Delta$ and a broad part around the phonon peak of the normal state
with the total spectral weight conserved.
Weight is transferred from the sharp peak to the broad peak at low
temperatures by a magnetic field.

As a model, we consider a simple ellipsoidal Fermi-surface,
$\epsilon_{\vec k} = \sum_i^d \frac{{\vec k}_i^2}{2m_i}$,
and calculate
$\Pi ({\vec Q},\omega)$ in the superconducting state for ${\vec Q}$
near the extremum vectors ${\vec Q_0}$.

We write the model Hamiltonian which describes a system
of electron interacting via a potential V,
\begin{equation}
H_{el}=\sum_{{\vec k}} \epsilon_{{\vec k}} \Psi_{{\vec k}}^{\dagger}
\tau_3 \Psi_{{\vec k}} + \frac{1}{2} \sum_{{\vec k},{\vec k^{\prime}},
{\vec Q}} V({\vec k},{\vec k^{\prime}},{\vec Q}) 
(\Psi_{{\vec k}+{\vec Q}}^{\dagger} \tau_3 \Psi_{{\vec k}})
(\Psi_{{\vec k^{\prime}}-{\vec Q}}^{\dagger} \tau_3 \Psi_{{\vec k}^{\prime}}),
\end{equation}
in Nambu notation\cite{nambu},
where the electron and annihilation operators are written as 
two-component vectors,
\begin{equation}
\Psi_{\vec k}=\left( \begin{array}{c} c_{{\vec k}\uparrow}\\
c_{-{\vec k} \downarrow}^{\dagger} \end{array}\right),
\Psi_{\vec k}^{\dagger}=(c_{{\vec k}\uparrow}^{\dagger},
c_{-{\vec k}\downarrow}),
\end{equation}
and the $\tau$'s are Pauli matrices.
The potential $V({\vec k},{\vec k^{\prime}},{\vec Q})$ includes 
the electron-electron Coulomb repulsion 
as well as the attractive interaction mediated by the phonons.
As usual\cite{nambu} we rewrite Eq. (3) as

\begin{equation}
H_{el}=H_0+H_1
\end{equation}
where $H_0$ is the BCS reduced Hamiltonian,
\begin{equation}
H_0= \sum_{{\vec k}} \Psi_{{\vec k}}^{\dagger}
(\epsilon_{{\vec k}} \tau_3+ \Delta \tau_1)  \Psi_{{\vec k}},
\end{equation}
and $H_1$ now includes the second term of Eq. (3) minus the electron-electron
effective interaction absorbed in Eq. (6).

The lowest order polarizability, Fig. 1 (a), at $T=0$ is
\begin{equation}
\Pi_0 ({\vec Q},\omega) = -i \int \frac{d^3 k d\omega^{\prime}}{(2 \pi)^4}
                Tr [\tau_3 G({\vec k}+{\vec Q},\omega+\omega^{\prime}) 
                   \tau_3 G({\vec k},\omega^{\prime})].
\end{equation}
where $G$ is the single particle Green's function for the BCS reduced 
Hamiltonian.

\begin{equation}
G({\vec k},\omega) = \frac{\omega I +\epsilon_{\vec k} \tau_3 +
\Delta \tau_1}{\omega^2 - E_{\vec k}^2 + i\delta},
\end{equation}
where $E_{\vec k}^2 =(\epsilon^2_{\vec k}+\Delta^2)$.

To evaluate Eq. (7), it is helpful, after performing the frequency integral, 
 to divide the
momentum space into parts in which $E_{\vec k}$ and $E_{{\vec k}+{\vec Q}}$
 are close to $\Delta$,
say between $\Delta$ and $s\Delta$, $s \approx O(2)$, 
and those in which they are  larger than $s\Delta$.
In the latter $E_{{\vec k},{\vec k}+{\vec Q}}$ may be replaced by 
$\epsilon_{{\vec k},{\vec k}+{\vec Q}}$,
so that the integrand is the same as in the normal state.
Any new features expected in the superconducting state can arise only 
from the singular density of one-particle states for energies near
$\Delta$, i.e., from regions of momentum space where both 
$E_{\vec k}$ and $E_{{\vec k}+{\vec Q}}$
are near $\Delta$.
The contribution of {\it this} region
in three dimensions is,

\begin{eqnarray}
{\rm Re} \Pi_0^{sc}({\vec Q},\omega) &\simeq&  
\frac{N(0) \Delta}
{E_F } sgn(\delta \omega)\hspace{1mm}  {\rm ln}\frac{E_F}{\sqrt{|\delta \omega| \Delta}}
\nonumber\\
{\rm Im} \Pi_0^{sc}({\vec Q},\omega)& \simeq & \cases { 0 & $ \delta \omega < 0 $\cr
-\frac{\pi}{2} \frac{ N(0) \Delta}
{E_F}
&  $\delta \omega > 0 $, \cr}
\end{eqnarray} 
where $\delta \omega \equiv \omega - 2 \Delta(1+\xi^2 q^2/8)$, 
$N(0) =v \frac{\sqrt{m_{i \perp} m_{j \perp} m_{\parallel} E_F}} 
{2 \pi^2}$ where $v$ is the volume of the unit cell,
and $q$ defined to be 
$({\vec Q}-{\vec Q}_0)_{\parallel}$ satisfies $q \xi << 1$.

In two dimensions, ${\rm Re}\Pi_0^{sc} ({\vec Q}_0,\omega)$ is
proportional to $|2\Delta-\omega|^{-\frac{1}{4}}$ while
in one dimension (where the treatment here is quite inadequate)
it is proportional to $|2\Delta-\omega|^{-\frac{1}{2}}$.
For non-extremum ${\vec Q}$ spanning the Fermi-surface,
${\rm Re} \Pi_0^{sc} ({\vec Q},\omega)$ is proportional
to ${\rm ln} |2\Delta-\omega|$ in two dimensions\cite{yaco}
and $|2\Delta-\omega|^{\frac{1}{2}}$ in three dimensions.

The contribution of the other regions differs from the normal state only
through a lower cut off at $s\Delta$ so that ${\rm ln}(E_F/\omega)$ in Eq. (1)
becomes ${\rm ln}[E_F/(\omega+s\Delta)]$.
Since we will be concerned with $\omega \approx O(2\Delta)$,
this change is not of consequence, so that
\begin{equation}
{\rm Re} \Pi_0 ({\vec Q},\omega) \simeq {\rm Re} \Pi_0^{sc} ({\vec Q},\omega)
+{\rm Re} \Pi_0^{n} ({\vec Q},\omega)
\end{equation}

In the random phase approximation, Fig. 1 (b),
\begin{equation}
\Pi ({\vec Q},\omega) = \Pi_0 ({\vec Q},\omega)/(1+V({\vec Q}) 
\Pi_0 ({\vec Q},\omega)),
\end{equation}
The direct electron-electron interaction is more important
in $V({\vec Q})$ in Eq. (11) compared to the phonon-induced interaction
because of the cut-off $E_F$ in the former compared to Debye frequency 
in the latter.
Vertex correction to Eq. (11) only change the magnitudes calculated here
 without affecting the singularities.\cite{footnote2}
This is unlike the case ${\vec Q} \rightarrow 0$.\cite{littlewood}

Since $\Pi_0^{sc} ({\vec Q},\omega)$ is singular for ${\vec Q}={\vec Q_0}$ and
$\omega = 2 \Delta$ while $\Pi_0^{n} ({\vec Q},\omega)$ is smooth
and much larger than $\Pi_0^{sc}$ for all $\omega$ except near $2\Delta$,
polarizability  is approximately given by 

\begin{eqnarray}
{\rm Re} \Pi ({\vec Q}_0,\omega) &\simeq& {\rm Re} \Pi^{sc} ({\vec Q}_0,\omega)  {\hspace{2cm}}
\omega \approx 2\Delta\nonumber\\
{\rm Re} \Pi ({\vec Q}_0,\omega) &\simeq& {\rm Re} \Pi^n ({\vec Q}_0,\omega)     {\hspace{2cm}}
{\rm elsewhere},
\end{eqnarray}
Here  $\Pi^n ({\vec Q}_0,\omega)$ is the normal state polarizability and
\begin{equation}
{\rm Re} \Pi^{sc} ({\vec Q}_0,\omega) \simeq 
\frac{-N(0)}{\beta+V {\rm Re} \Pi_0^{sc}({\vec Q}_0,\omega)}.
\end{equation}
In Eq. (11), we have used ${\rm Re} \Pi^n_0 \simeq -N(0)$ and
$\beta \simeq 1-V N(0)$.
In three dimensions $\Pi^{sc} ({\vec q},\omega)$ always has a pole at
\begin{equation}
\omega=2\Delta(1-e^{-\frac{\beta}{\gamma}}),
\end{equation}
where $\gamma=\Delta N(0) V/E_F << 1$.
Note that this pole is very close to $2\Delta$ and has weight
\begin{equation}
r \simeq 2 E_F  e^{-\frac{\beta}{\gamma}}/V \hspace{1cm}   << 1.
\end{equation}

The phonons couple to the electronic charge density,
\begin{equation}
H_{el-ph}=\sum_{\vec k} g_Q (b_{\vec Q}+b_{-{\vec Q}}^{\dagger}) 
\Psi_{{\vec k}+{\vec Q}}^{\dagger} \tau_3 \Psi_{\vec k},
\end{equation}
where $g_Q$ is the electron-phonon coupling constant.
The phonon spectrum in a metal is described by
\begin{equation}
D^{-1}({\vec Q},\omega)=D_0^{-1}({\vec Q}, \omega)-\Sigma({\vec Q}, \omega),
\end{equation}
where $\Sigma({\vec Q}, \omega)=g_Q^2 \Pi({\vec Q},\omega)$ and
$D_0 ({\vec Q}, \omega)$ 
contains no renormalization due to electron-phonon interactions.
Using Eqs. (12) and (13) in Eq. (17),
\begin{eqnarray}
D ({\vec Q}_0, \omega) & = &
[\omega-\omega_{Q_0}-\frac{g_{Q_0}^2 r}{\omega-2\Delta}
+i \Gamma_{Q_0} \theta(\omega-2\Delta)]^{-1} \nonumber\\
& & +[\omega+\omega_{Q_0}+\frac{g_{Q_0}^2 r}{\omega+2\Delta}
-i \Gamma_{Q_0} \{1-\theta(\omega+2\Delta)\}]^{-1}
\end{eqnarray}
where $\omega_{Q_0}$ and $\Gamma_{Q_0}$ are the phonon frequencies 
and linewidths of
the normal state obtained from the normal contribution $\Pi^n ({\vec Q}_0,
\omega)$ to $\Pi ({\vec Q}_0, \omega)$.
The $\theta$-function in Eq. (18) takes into account that 
${\rm Im} \Pi ({\vec Q}_0, \omega)=0$ for $\omega < 2\Delta$.
Actually, we should include in Eq. (18) also a contribution to 
$\Gamma_{Q_0}$ due to ${\rm Im} \Pi^{sc} ({\vec Q}_0,\omega)$ which
increases damping of the normal state
if $\omega_{Q_0}$ is between $2\Delta$ and $\approx O(4 \Delta)$.

The electron-phonon coupling function
$g_{Q_0}^2 \simeq \lambda {\bar \omega} E_F$ where ${\bar \omega}$
is the order of the Debye frequency
and $\lambda$ is the dimensionless electron-phonon
coupling constant, typically  of $O(1)$.
Then $(g_{Q_0}^2 r)^{1/2}$ is typically much less than $2\Delta$.
In that case and for $\omega_{Q_0} > 2\Delta$,
Eq. (18) predicts a pole at
\begin{equation}
\omega \simeq 2\Delta-\frac{1}{2} g_{Q_0}^2 r/(\omega_{Q_0}-2\Delta) \equiv \nu,
\end{equation}
and peak with width $\Gamma_{Q_0}$ at
\begin{equation}
\omega \simeq \omega_{Q_0}+\frac{1}{2} g_{Q_0}^2 r/(\omega_{Q_0}-2\Delta).
\end{equation}
The spectral function of the pole near $2\Delta$ is
\begin{equation}
S({\vec Q}_0,\omega) \simeq \frac{g_{Q_0}^2 r}{4 (\omega_{Q_0}-2\Delta)^2
+g_{Q_0}^2 r} \delta (\omega-\nu),
\end{equation}
with the rest in the peak near $\omega_{Q_0}$.
If $\omega_{Q_0}$ is close to $2\Delta$, the pole is at 
$2\Delta-g_{Q_0} r^{1/2}$ and shares weight equally with the peak.
The spectral function, 
$S({\vec Q}_0,\omega)=-\frac{1}{\pi} {\rm Im} D ({\vec Q}_0,\omega)$ 
is illustrated in Fig. 2.
For $q \ne 0$, the new mode increases in energy by $O(\Delta
\xi^2 q^2/4)$, and is overdamped for $ q > \xi^{-1}$.

In comparing with experimental results\cite{kawano,stassis}
in $(Lu,Y)Ni_2B_2C$, one should
bear in mind that the $d=3$ results in normal state  are explained
in a detailed calculation\cite{weber}
of $\Pi_0({\vec Q}, \omega)$ based on the actual complicated 
band structure and phonon renormalizations using electron-phonon
coupling incorporating local filed effects given by methods of 
Ref. ~\cite{varma}.
We have studied only the effects due to superconductivity assuming 
$\omega_{Q_0}$ in Eq. (18) to be the observed normal state value 
just above $T_c$.
The band structure of $(Y,Lu)Ni_2B_2C$ shows the extremum vectors
in the $(a^*,b^*)$ plane for ${\vec Q}_0 \sim (0.5,0,0)$,
and weak dependence on momentum in the $c^*$ direction.\cite{rhee}
The local field effects quite generally enhance the magnitude of the
effects in the superconducting state just as they do in the normal state.
But since they are not included in our calculation not too much reliance
can be put on the estimate of magnitude of the coefficients in
Eqs. (19)-(21).
Moreover band structure calculations shows significant nesting near
${\vec Q}_0$ of the anomaly.
For well nested surfaces, the $d=2$ result quoted below Eq. (9) 
is relevant and the effects we discuss are correspondingly stronger.

$YNi_2B_2C$ is a particularly good candidate for the effects
discussed here because the transverse acoustic phonon at
${\vec Q}_0 \simeq (0.525,0,8)$ softens to an energy $\simeq 7$ meV just above
$T_c$, not too far above $2\Delta \simeq 4.3$ meV.
Thus the weight transferred to the peak near $2\Delta$ is significant 
and estimated from Eq. (21) to be the experimental value\cite{kawano}
$\approx 0.3$ if  $r/2\Delta=0.4 N(0)$
and $\gamma \approx O(0.01)$ at $T=0$.
Actually, we have no independent way of estimating $r$.
This value is reasonable only if $\beta \approx O(0.1)$,
which implies a phonon close to an instability in the normal phase,
as in these compounds.

A simple extension of the calculation above shows  
that as the temperature is decreased
and $\Delta(T)$ grows, weight is transferred to the sharp peak at the expense of
the broad peak with the total weight constant.
Similarly a magnetic field depresses the weight and the frequency of 
the sharp peak.
These are as observed\cite{kawano}.
The theory would also predict a temperature dependence to the frequency 
of the peak $\simeq 2\Delta(T)$.
In the experiment\cite{kawano,stassis}, the frequency of the new peak appears
not to follow the BCS dependence of $2\Delta(T)$.
This may be so because of strong-coupling effects.
Note also that the relative intensity of the peak drops
rapidly as its frequency decreases below $2\Delta(0)$, i.e.,
near $T_c$, making it harder to observe.

We finally comment on other possible explanation of the data.
The observed peaks cannot be the large ${\vec Q}$-extensions of 
the amplitude modes of 
superconductors\cite{littlewood} near ${\vec Q} \simeq 0$ observed in Raman
scattering\cite{soor}
since such modes are heavily damped for $|{\vec Q}| \ge \xi^{-1}$.

Another possibility is that the peak is just the narrowing of the linewidth
as the phonon frequency goes below $2\Delta$.
This would predict that the number of peaks for a fixed ${\vec Q}_0$
as a function of $\omega$ remains the same in the normal state and
the superconducting state.
In Kawano et al's data\cite{kawano}, an extra new peak for $T<T_c$ 
clearly appears, as in the theory here.
In the data of Stassis et al in $LuNi_2B_2C$\cite{stassis}, 
an extra peak is not visible.
In this compound the phonon frequency of the soft transverse acoustic
mode in the normal state just above $T_c$ is lower than that in
$Y$-based compound, and in fact close to the sharp peak appearing
well below $T_c$. Recall that if $\omega_{Q_0}$ is
close to $2\Delta(0)$,
the two peaks in the superconducting state are split only by
$g_{Q_0} r^{1/2}$ as discussed below Eq. (21).
The two peaks may then lie within the experimental resolution
of about 1 $meV$.
Further experiments are required to elucidate the situation.

An interesting conclusion  from  our work is that
a second order transition to a charge density wave state at 
a temperature, $T_{CDW}$, below the superconducting transition
temperature, $T_c$, may not be allowed.
A second order transition requires $\omega_{Q_0}(T) \rightarrow 0$ as
$T \rightarrow T_{CDW}$.
But $\omega_{Q_0} (T)$ is bounded from below by 
$2\Delta(T)$ for $T_c > T_{CDW}$,
because of the singular contribution to the phonon self-energy as 
$\omega_{Q_0}$ approaches $2\Delta(T)$.
The variation of the phonon frequencies as a function of temperature
is schematically illustrated in Fig. 3.
A first order transition is, of course, allowed.
These matters need further investigation. 

In summary, we find the singularities in the polarizability of 
s-wave superconductor at nesting vectors in three and two dimensions,
which manifest themselves as poles in the lattice response function
at $\omega$ just below $2\Delta$ with its weight dependent on the frequency 
of the normal state phonon at the nesting vector.
These results are in general accord with observations.
A straightforward extension of these results can be made to d-wave 
superconductors. If the maximum of the gap function in such 
superconductors is along the nesting direction,
the results for d-wave superconductors are expected to be 
similar to those obtained here.

We are grateful to Dr. H. Kawano and Prof. C. Stassis for discussions
of the experimental results and to Prof. B. Altshuler for pointing
out an error in the calculation.
We have also learned that Dr. S. K. Sinha(private communication)
has arrived at conclusions similar to those in the present paper.
One of authors (H. Y. K.) thanks E. Abrahams, S.-W. Cheong,
Y. B. Kim, and P. B. Littlewood for helpful discussions.
This work is supported in part by the Korea Research Foundation 
and NSF Grant No. DMR-96-32294 (H. Y. K.).

\newpage
{\large Figure Captions}\\
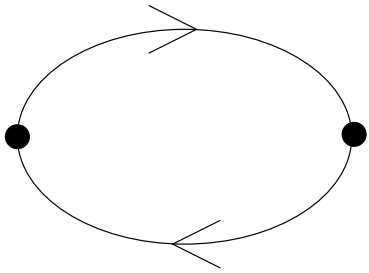
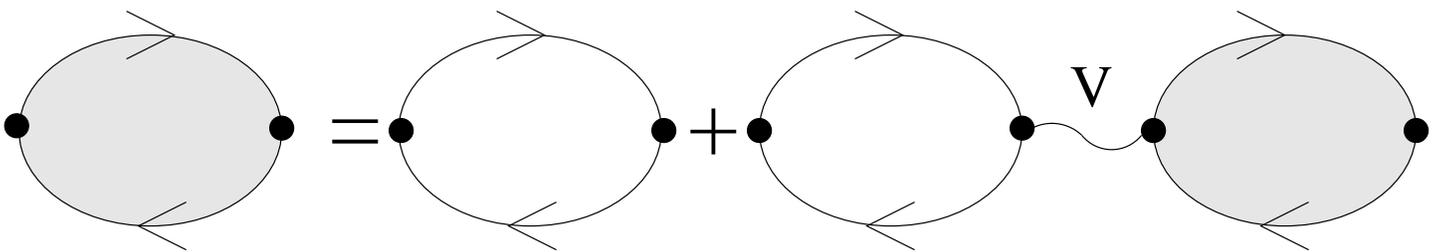
\noindent Fig. 1: (a) Lowest order of polarizability $\Pi_0({\vec Q},\omega)$,
and (b) Polarizability $\Pi({\vec Q},\omega)$ in random phase approximation.
\\

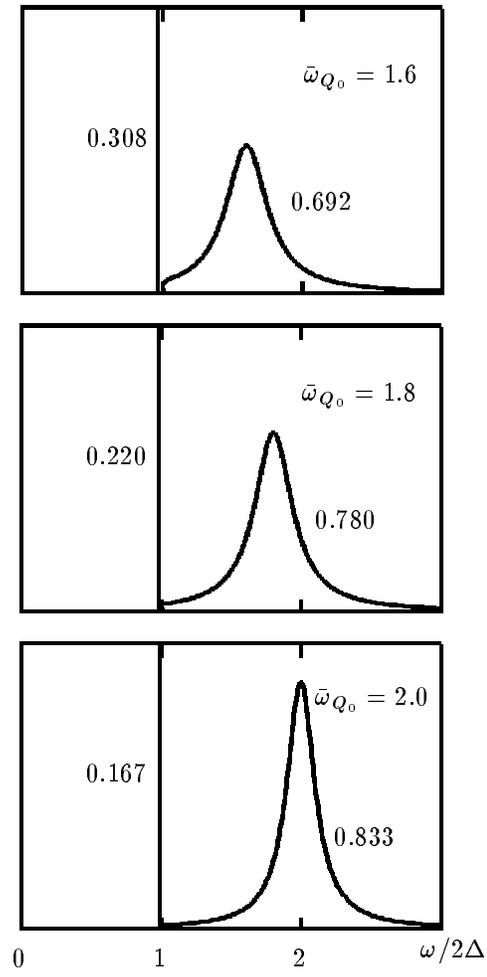
\noindent Fig. 2: Phonon spectral function $S({\vec Q}_0,\omega)$ 
calculated from Eq. (21) for various
${\bar \omega_{Q_0}}$ where ${\bar \omega_{Q_0}}=\omega_{Q_0}/2\Delta$,
and $r/2\Delta=0.4 N(0)$.
\\

\noindent Fig.3: Schematic behavior of the phonons as a function
of temperature.
For $T_c$ below the charge density wave transition temperature
$T_{CDW}$, there is only a single branch $\omega_{Q_0} \rightarrow 0$ 
as $T \rightarrow T_{CDW}$,
as shown by the dashed line.
For $T_c > T_{CDW}$, the soft phonon branch splits into two for $T < T_c$,
the lower one $\omega_1$ is slightly below $2\Delta(T)$ and the upper one, 
$\omega_2$ considerably broadened with its lower edge at 
slightly above $2\Delta(T)$.


\begin{references}

\bibitem{kohn} W. Kohn, Phys. Rev. Lett. {\bf 2}, 393 (1959).
\bibitem{varma} C. M. Varma and W. Weber, Phys. Rev. B {\bf 19}, 6142
(1979); C. M. Varma and A. L. Simons, Phys. Rev. Lett. {\bf 51}, 138 (1983).
\bibitem{2kpol} A. L. Fetter and J. D. Walecka, 
{\it Quantum Theory of Many Particle Systems}
(McGraw Hill, New York, 1971).
\bibitem{kawano} H. Kawano et. al, Phys. Rev. Lett. {\bf 77}, 4628 (1996).
\bibitem{stassis} C. Stassis st. al, unpublished.
\bibitem{nambu} Y, Nambu, Phys. Rev. {\bf 117}, 648 (1960);
J. R. Schrieffer, {\it Theory of Superconductivity}
(Benjamin/Cummings, Reading, Massachusetts, 1964).
\bibitem{yaco} I. I. Mazin and V. M. Yakovenko, Phys. Rev. Lett.
 {\bf 75}, 4134 (1995).
%\bibitem{footnote} Because of simple ellipsoidal geometry of the assumed 
%Fermi-surface, 
%the coefficient of the singular term does not depend on whether the direction
%of ${\vec Q}$, spanning vector, across the major or the minor axis. 
%However, if $Q$ deviates from these axis in such
%a way that we can not expand ${\vec Q}$ near ${\vec Q}_0$, an anisotory
%with respect to $x$ and $y$ should occur.
%%This is in general not true.
\bibitem{footnote2}
As has been noted before, $\Pi_0({\vec Q},\omega)$ for ${\vec Q} \simeq 0$
has an inverse square root singularity at $\omega=2\Delta$
in all dimensions\cite{littlewood,balse}.
But this singularity is removed in a gauge-invariant theory
using Coulomb interaction for $V({\vec Q})$\cite{littlewood}.
By contrast $V({\vec Q})$ for large ${\vec Q}$ may be taken to be a constant
and only shifts the position of the pole in Eq. (12),
so that the vertex correction does not affect the singularity
in this case.
\bibitem{littlewood} P. B. Littlewood and C. M. Varma, Phys. 
Rev. Lett. {\bf 47}, 811 (1981); Phys. Rev. B {\bf 26}, 4883 (1982).
\bibitem{balse} C. A. Balseiro and L. M. Falicov, Phys. Rev. Lett. {\bf 45},
662 (1980).
\bibitem{weber} W. Weber, unpublished.
%\bibitem{siegrist} T. Siegriest, {\it et al}, Nature, {\bf 367}, 254 (1994).
\bibitem{rhee} J. Y. Rhee, X. Wang, and B. N. Harmon, Phys. Rev. B
{\bf 51}, 15585 (1995).
\bibitem{soor} R. Sooryakumar and M. V. Klein, Phys. Rev. Lett. {\bf 45},
660 (1980).
%\bibitem{axe} J. D. Axe and G. Shirane, Phys. Rev. B {\bf 8}, 
%1965 (1973).

\end {references}
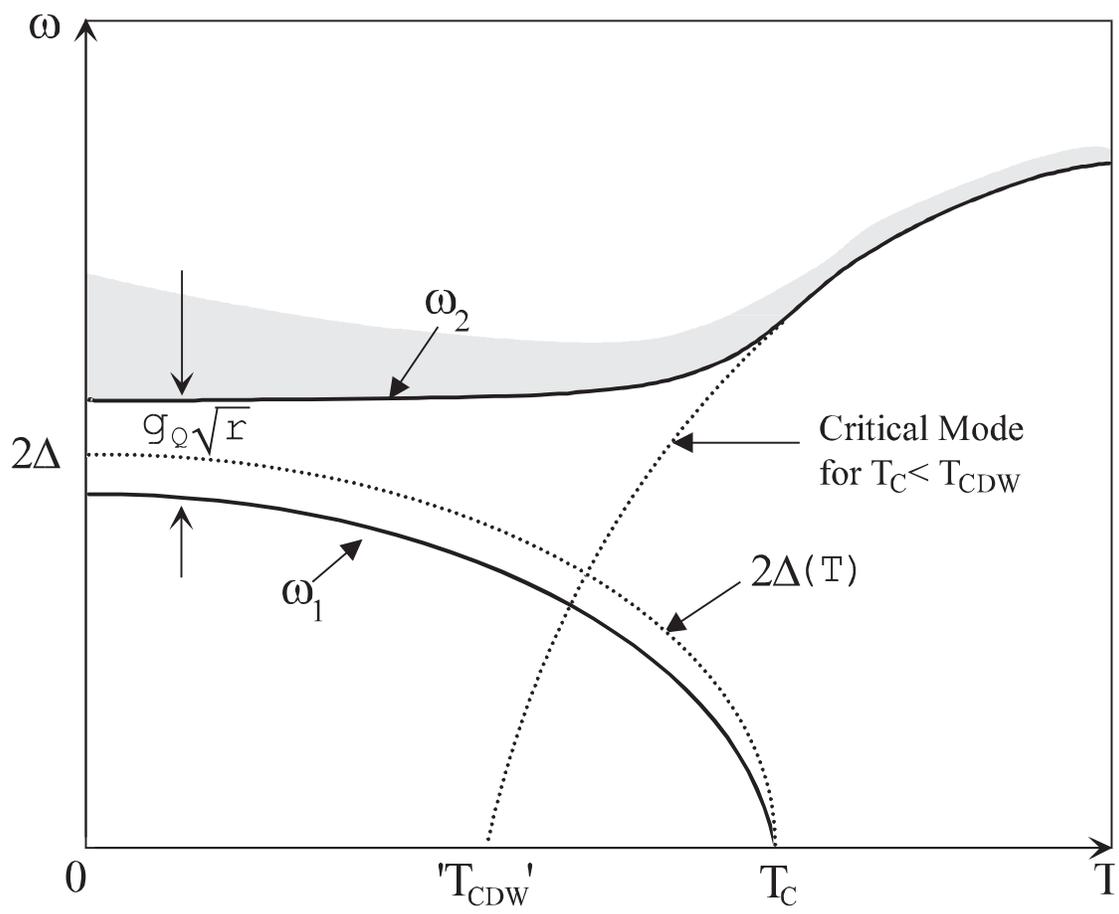
\end{document}